\documentclass[useAMS,usenatbib]{mn2e}
\usepackage{epsfig,amsmath,natbib}
\usepackage{color}

\def\be{\begin{equation}} 
\def\ee{\end{equation}} 
\def\ba{\begin{eqnarray}} 
\def\ea{\end{eqnarray}}

\def\msun{{\Msun}}

\def\HH{${\rm {H_2}}\,\,$}

\def\fesc{f_{\esc}} 
 
\def\HII{\hbox{H~$\scriptstyle\rm II\ $}}

\def\gsim{\lower.5ex\hbox{\gtsima}} 
\def\lsim{\lower.5ex\hbox{\ltsima}} \def\gtsima{$\; \buildrel > \over 
\sim \;$} \def\ltsima{$\; \buildrel < \over \sim \;$} \def\prosima{$\; 
\buildrel \propto \over \sim \;$} \def\gsim{\lower.5ex\hbox{\gtsima}} 
\def\lsim{\lower.5ex\hbox{\ltsima}} 
\def\simgt{\lower.5ex\hbox{\gtsima}} 
\def\simlt{\lower.5ex\hbox{\ltsima}} 
\def\simpr{\lower.5ex\hbox{\prosima}} \def\la{\lsim} \def\ga{\gsim} 
  
 \def\gtsima{$\; \buildrel > \over \sim \;$} 
\def\ltsima{$\; \buildrel < \over \sim \;$} 
\def\gsim{\lower.5ex\hbox{\gtsima}} 
\def\lsim{\lower.5ex\hbox{\ltsima}} 
\def\simgt{\lower.5ex\hbox{\gtsima}} 
\def\simlt{\lower.5ex\hbox{\ltsima}} 
\def\simpr{\lower.5ex\hbox{\prosima}} 
\def\la{\lsim} 
\def\ga{\gsim} 
\def\de{{\rm d}}

\def\fesc{f_{esc}}

\def\msun{\,{\rm \Msun}}

\def\E3{{\cal E}_{\rm g}^{III}}

\def\Msun{\rm M_\odot}

\def\rvir{r_{vir}}
\def\rvir{r_{vir}}

\def\Tvir{T_{vir}} 
\def\r12{r_{1/2}} 
\def\x12{x_{1/2}} 
\def\v12{v_{1/2}}

\def\Msf{M_{sf}}
 
\title[Escape fraction of ionizing radiation]{Escape Fraction of
Ionizing Radiation from Starburst Galaxies at High Redshifts}

\author[Ferrara \& Loeb]{Andrea Ferrara$^{1}$ \& Abraham Loeb$^{2}$\\
$^{1}$ Scuola Normale Superiore, Piazza dei Cavalieri 7, I-56126 Pisa,
Italy\\ $^{2}$ Astronomy department, Harvard University, 60 Garden
Street, Cambridge MA 02138, USA\\ }

\begin{document} 
 
\date{\today} 
 
\pagerange{\pageref{firstpage}--\pageref{lastpage}} \pubyear{2012}
 
\maketitle 
 
\label{firstpage} 
\begin{abstract} 
Recent data indicates that the cosmic UV emissivity decreased with
decreasing redshift $z$ near the end of reionization. Lacking evidence
for very massive early stars, this could signal a decline with time
in the mass-averaged escape fraction of ionizing radiation from
galaxies $\langle \fesc \rangle$ at $z\simgt 6$.  We calculate the
evolution of ionization fronts in dark matter halos which host gas in
hydrostatic equilibrium at its cooling temperature floor ($T \approx
10^4$ K for atomic hydrogen).  We find a high escape fraction only for
the lowest mass halos (with $M< 10^{8.7} \msun$ at $(1+z)=10$)
provided their star formation efficiency $f_\star \simgt 10^{-3}$.
Since the low-mass galaxy population is depleted by radiative
feedback, we find that indeed $\langle \fesc \rangle$ decreases with
time during reionization.
\end{abstract}


\section{Introduction}
\label{Mot}

The first generation of galaxies is expected to have reionized cosmic
hydrogen by a redshift $z\approx 7$ \citep{Loeb13}. One of the most
important unknown parameters regulating the way in which reionization
proceeds is the fraction of ionizing photons, $\fesc$, that escape
outside the virial radius of the galaxies in which they were produced
and into the surrounding intergalactic medium (IGM).  The growth of
ionized regions with cosmic time depends on the average value of this
parameter over viewing angle per galaxy and over the galaxy population
\citep{Wyithe03, Trac07, Bouwens12, Finkelstein12}.

Observations of galaxies at $z= 1-3$ indicate a broad set of escape
fraction values ranging from a few percent to tens of percent
\citep{Steidel01, Fernandez03, Shapley06, Inoue06, Siana07,
Giallongo08, Iwata09, Boutsia11, Grazian11, Vanzella12a,
Vanzella12b}. This potentially reflects strong variations of the
escape fraction with viewing angle and evolutionary time.

A calculation of the escape fraction of ionizing radiation from first
principles is difficult, as it depends on the spatial distribution of
ionizing sources relative to the neutral hydrogen in galaxies and is
therefore sensitive to the small scale clustering of young stars and
the interstellar medium.  Due to their higher density, high-redshift
galactic disks are expected to have small escape fractions during
reionization \citep{Wood00,Gnedin08, Razoumov06, Wise09}. Even for
local disk galaxies, \citet{Dove00} concluded that $\fesc$ should not
exceed a few percent in most cases, as a result of efficient radiation
trapping by the shells of the expanding superbubbles around OB star
associations.

Since the mean density of the Universe (and hence galaxies) scales as
$(1+z)^3$, theoretical calculations tend to conclude that $f_{\rm
esc}$ should decrease with increasing redshift.  This expectation
appears to be in conflict with recent reionization models and data
\citep{Bolton07, Salvaterra11, Mitra12a, Mitra12b, Kuhlen12,
Finkelstein12} that point towards the need for an increasing UV
emissivity towards high redshifts. This can be achieved in two
ways. The first is to postulate an increasingly top-heavy initial mass
function (IMF) of stars that would increase the number of photons
emitted per baryon incorporated in stars.  Although possible, this
explanation seems to be disfavored by the lack of observational
evidence for an early population of substantially more massive stars
\citep{Cayrel04, Caffau11} and by recent numerical simulations
\citep{Greif12}. A more appealing solution is represented by a
possible increase in the average escape fraction $\langle f_{\rm
esc}\rangle$ during reionization, recently suggested by \cite{Mitra12b}.  
Here we explore a novel physical explanation for this unexpected trend.

We show that the lowest-mass galaxies near or below (so-called
mini-halos) the hydrogen cooling threshold have high escape fractions
but their contribution to the UV production rate decreases with cosmic
time due to large-scale radiative feedback processes during
reionization that either photo-heat or sterilize them by dissociating
their H$_2$.  However, internal feedback is of foremost importance as
well: as these small systems achieved large $\fesc$ by rapidly (within
$\sim 10^4$ yr) ionizing their entire interstellar medium, this gas
will become loosely bound to the host galaxy. Under these conditions
it is difficult to sustain a continuous mode of star formation (SF),
particularly because following the death of massive stars powerful
supernova explosions will clear the gas out of the potential well
before a stable disk structure is established. More massive galaxies
will form a disk with the net result of trapping most of their
ionizing photons.

If high-redshift dwarf galaxies form stars predominantly in episodic
bursts, as suggested by the high incidence of galaxy mergers at
increasing redshift \citep{Munoz11}, the conventional argument that
their star formation efficiency is suppressed by supernova (SN)
feedback may not be valid. In particular, SN feedback is inhibited if
the duration of the starburst is shorter than $\sim 10~{\rm Myr}$, the
lifetime of SN progenitors. In this regime, the global radiative
feedback sets the amount of gas initially available for making stars
and hence the overall efficiency of star formation, $f_\star$.  In our
calculations, we will calibrate the value of $f_\star$ based on recent
measurements of the cosmic stellar mass density at high redshifts
\citep{Gonzalez11}.

For an instanteneous starburst, $f_{\rm esc}$ can be large only if an
escape route for ionizing photons is opened within a few Myr, prior to
the death of the massive stars that produce these photons. This
process cannot be mediated by supernova explosions, which occur after
the emission of ionizing photons has already started to decline.  The
key question is therefore whether the surrounding blanket of absorbing
hydrogen atoms can be ionized before the massive stars in the
starburst end their life. We address this question within the context
of a simple model in which we populate dark matter halos with
primordial gas that cooled to the temperature floor of atomic
hydrogen, $\sim 10^4$K.

Our basic point is simple: massive galaxies exist at all redshifts and
the escape fraction of ionizing radiation from them is calculated and
observed to be small; however, the lowest-mass galaxies ($\Tvir \simlt
10^4$ K) exist only before and during reionization and have a high
$\fesc$.  Our assumption of a short-lived star formation phase without
SN feedback is self-consistent, as we will show that photoionization
heating of the gas suppresses star formation in dwarf galaxies after a
period of time much shorter than the lifetime of massive stars.

In \S 2 we describe the details of our calculation and in \S 3 we
present our numerical results. Finally, we discuss our main
conclusions in \S 4.  Throughout the paper, we adopt the WMAP7 set of
cosmological parameters with $\Omega_m = 0.27$, $\Omega_{\Lambda} = 1
- \Omega_m$, $\Omega_b h^2 = 0.023$, $h=0.71$, $\sigma_8=0.81$, $
n_s=0.97$, and $\de n_s/\de \ln k =0$ \citep{Larson11}.

\section{Method of Calculation}
\label{Met}

We derive the escape faction of the ionizing photons, $\fesc$, based
on a set of simplifying assumptions.  We consider a spherical dark matter
halo of mass, $M$, virializing at redshift $z$ and characterized by an
internal density (spherically averaged) profile following the
Navarro-Frenk-White (\cite{Navarro97}, NFW) form, 
\be 
\rho(r) = \frac{\rho_c
\delta_c }{cx(1+cx)^2},
\label{eq1}
\ee 
where $x \equiv r/\rvir$, $\rvir$ is the virial radius of the
system, $c$ is the halo concentration parameter, $\delta_c =
(18\pi^2/3)c^3/F(c)$ is a characteristic overdensity, and \be F(t) =
\ln(1+t) - \frac{t}{1+t}.
\label{eq2}
\ee For the concentration parameter we use the results of
\cite{Prada12}, extrapolating them towards lower masses and higher
redshift when necessary. The circular velocity, $v_c^2(r) = GM(r)/r$,
can be expressed as \be v_c^2 = V_c^2 \frac{F(cx)}{xF(c)},
\label{eq3}
\ee with $V_c^2 \equiv GM/\rvir$. Baryons fall into the dark matter
halo potential well are shock-heated to the virial temperature $\Tvir
= (\mu m_p/2k) V_c^2$, and then relax to a hydrostatic configuration
\citep{Makino98}, \be \ln\rho(r) = \ln \rho_0 -
\frac{V^2}{V_c^2}[v_e^2(r=0)-v_e^2(r)].
\label{eq4}
\ee
%
%
\begin{table}
\begin{minipage}{85mm}
\caption{Values of central number density, $n_0$ (in units of cm$^{-3}$)  for dark matter halos of different 
total mass, $M$, at $z=9$ and for different values of the gas temperature, $T$, expressed in terms of 
the virial temperature $\Tvir$ through the parameter $V^2=\Tvir/T$.}
\begin{center}
\begin{tabular}{|c|c|c|c|c|}
\hline\hline
$\log_{10}\Tvir$ [K]  &             4.31     &         4.98        &         5.65             &          6.32                  \\ \hline
$M$ [$\msun$]      &      $10^8$ &      $10^9$ &     $10^{10}$   & $10^{11}$            \\ \hline 
$V^2=1.0$                &   20.6  &         23.9         &         36.0        &          167.8             \\ \hline 
$T =2\times 10^4$ K   &   24.3          &        $1.9\times 10^4$         &       $4.3\times 10^6$                 &       ---                    \\ \hline 
\end{tabular}
\end{center}
\label{Tab1}
\end{minipage}
\end{table}
Here, $v_e$ is the halo escape velocity from the halo, 
\be v_e^2(r) = 2 V_c^2 \frac{F(cx)+
cx/(1+cx)}{xF(c)};
\label{eq5}
\ee $\mu=1.22$ is the mean molecular weight of a neutral primordial
H-He gas; $k_B$ is the Boltzmann constant and $m_p$ the proton
mass. We have also inserted in equation (\ref{eq4}) the extra factor
$V^2=V_c^2/c_s^2 = \Tvir/T > 1$, where $c_s^2 = 2k_BT/\mu m_p$ is the
effective sound speed at a generic temperature $T\neq \Tvir$. The
standard case in which $T=\Tvir$ is obtained by setting $V=1$. By
manipulating equation (\ref{eq4}) we then get the final expression for
the gas density profile as a function of radius, \be \rho(x) =\rho_0
e^{-\gamma^2 V^2} (1+cx)^{\gamma^2 V^2/cx} \equiv \mu m_p n_0 \eta(x);
\label{eq6}
\ee where we have defined $\gamma^2=2c/F(c)$, and the ancillary
function $\eta(x) = e^{-\gamma^2 V^2} (1+cx)^{\gamma^2 V^2/cx}$.  The
central density, $\rho_0$, is obtained by requiring that the total gas
mass in the halo is equal to the cosmological value, i.e. $M_b =
(\Omega_b/\Omega_m) M = f_b M$. This procedure yields \be \rho_0 =
(18\pi^2/3) f_b c^3 e^{\gamma^2 V^2} \left[ \int_0^c dt
(1+t)^{\gamma^2 V^2/t} t^2\right]^{-1} ,
\label{eq7}
\ee Table 1 provides the values of the central number density, $n_0$,
for different halo masses and $V$ values at a fixed redshift,
$z=9$. As expected, $n_0$ increases in larger masses even in the case
$V=1$ in which the gas is at the virial temperature of the halo. This
reflects the strong dependence of $n_0$ on the concentration
parameter, which increases from $c\approx 4$ for $M=10^8 \msun$ to
$c\approx 6.5$ for $M=10^{11} \msun$. This increasing trend is
amplified if the gas is cooler than the virial temperature ($V>1$) or
if it is thermostated at a fixed temperature (such as $T=2\times 10^4$
K in Table 1), leading to unrealistic densities for the most massive
halos with $M=10^{11} \msun$ (which represent rare $3\sigma$
fluctuations of the density field at $z=9$). For massive halos we
obtain artificially high densities because we ignore angular
momentum. Once the gas condenses to a sufficiently small scale, its
rotation will halt contraction and it will settle to a disk. Disks
were considered by \cite{Wood00} and \cite{Dove00} who independently
concluded that the escape fraction was negligible for a smooth gas
distribution in high-redshift disks and also local spirals. We
therefore argue that once the gas contracts enough to make a disk, its
density will not increase by as much as our spherical model predicts
but UV photons will not be able to escape from it anyway, based on
earlier studies.

\subsection{Ionization front evolution}
Having specified the gas density distribution inside halos, we may now
derive the evolution of an ionization front (IF) driven by the
emission of photons with energy $>13.6$eV from a burst of star
formation at the halo center.  We assume that a fraction $f_\star$ of
the available gas, $f_b M$, of a given halo is instantaneously
converted into stars distributed in mass according to a Salpeter IMF
in the range ($m_{low}, m_{up}$) = ($1 \msun, 100 \msun$) and
with absolute metallicity $Z=10^{-3}$. 
The ionizing photon production rate, $Q(t)$, by the stellar cluster
can be computed exactly from population synthesis models: we use here
\texttt{Starburst99}\footnote{http://www.stsci.edu/science/starburst99/}
\cite{Leitherer99}. The time dependence of the production rate of
ionizing photons rate under these conditions is, 
\be Q(t) =
\frac{Q_0}{1+(t/t_0)^4} ,
\label{eq8}
\ee with ($Q_0, t_0$) = ($10^{47} \rm s^{-1} \msun^{-1}, 10^{6.6}$
yr). Equation (\ref{eq8}) illustrates the important point that after
$\sim 4$ Myr, the production rate of ionizing photons rapidly drops
as a result of the death of short-lived massive stars. This has
important implications for $\fesc$ as discussed hereafter. Note that
equation (\ref{eq8}) implies that the number of ionizing photons
emitted per baryon incorporated into stars is $N_\gamma \approx 0.5 Q
t_0 m_p/\msun =5\times 10^3$.

The time evolution of the IF radius, $r_I$, is described by an
ordinary differential equation that expresses a detailed balance between
the ionization and recombination rates within the volume enclosed by the
IF, \be \frac{dr_I}{dt} = \frac{1}{4\pi n_H r_I^2} \left[ Q(t) -
\frac{4\pi}{3}r_I^3 n_H^2 \alpha^{(2)}\right],
\label{eq9}
\ee where $n_H=0.92 n$ is the hydrogen density for a primordial gas
and $\alpha^{(2)} = 2.6\times 10^{-13} (T/10^4{\rm K})^{-1/2}$ is the
Case B recombination rate of hydrogen \citep{Maselli03}. Normalizing
by an effective recombination rate in the halo, $ {\cal{\dot N}}_{rec}
= (4\pi/3) \alpha^{(2)} n_{H0}^2 \rvir^3$, and adopting a
dimensionless time variable $\tau=t/t_{rec}=\alpha^{(2)} n_{H0} t$,
equation (\ref{eq9}) can be written in a dimensionless form,
\be \frac{dx}{d\tau} = \frac{Q(t)}{3 {\cal{\dot N}}_{rec}
\eta(x) x^2} - \frac{1}{3} x \eta(x).
\label{eq10}
\ee Equation (\ref{eq10}) describes the expansion of the \HII region
into the stratified gas distribution within the halo. As the density
decreases outwards, the IF accelerates and eventually exits the virial
radius at a time $t_{out} \equiv t(x=1)$.  Until $t_{out}$ all the
ionizing photons are absorbed inside the halo\footnote{Here we ignore
the possibility of a highly inhomogeneous medium in which low density
channels for escape exist.}; hence, $\fesc\approx 0$.  However, for $t
> t_{out}$, ionizing photons will be only used \emph{to keep} the halo
ionized by balancing recombinations within it, but a large fraction of
them will be able to escape into the IGM, thus making $\fesc>0$. We
then express the net escape fraction as follows: \be \fesc=
\frac{\int_{\tau(x=1)}^\infty d\tau Q(t)}{\int_{0}^\infty d\tau Q(t)}
.
\label{eq11}
\ee As we will see shortly, though, either the IF is efficiently
confined by the density, ${dx}/{d\tau} \rightarrow 0$, or the blow out
will occur on a very short time scale thus making $\fesc \approx 1$.
We will refer to these two different situations as a ``confined'' or
``unconfined'' IF in the rest of the paper.

\section{Results}
\label{Res}
\begin{figure}
\includegraphics[width=90mm]{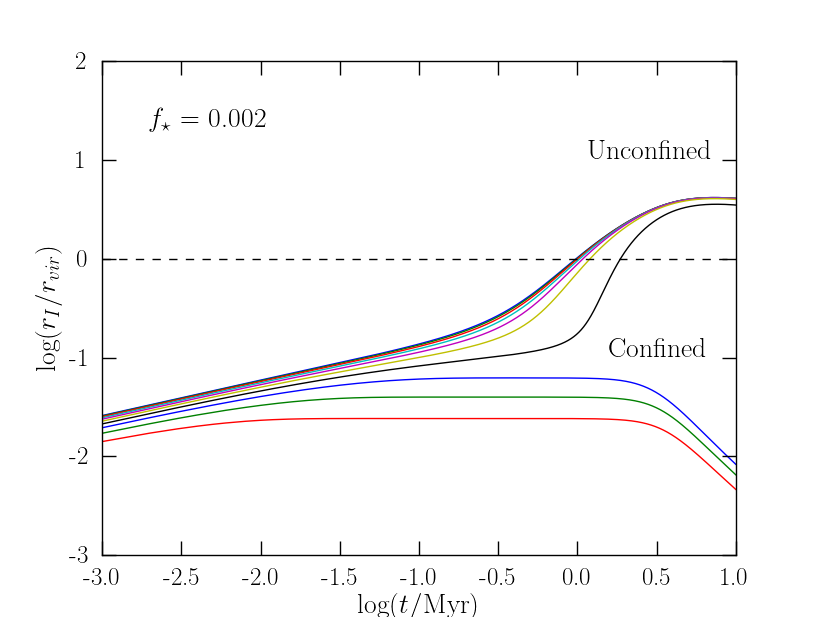}
\caption{Time evolution of the ionization front radius, $r_I$, at
$z=9$ within halos of different total mass $M=10^{8.+j/3} \msun$, with
$j=0,..,9$ from the top curve ($M=10^8 \msun$) to the bottom one
($M=10^{11} \msun$).  The assumed star formation efficiency is
$f_\star=0.002$ and the gas temperature factor $V=({\Tvir/T
})^{1/2}=1$.  }
\label{Fig1}
\end{figure}

\begin{figure}
\includegraphics[width=90mm]{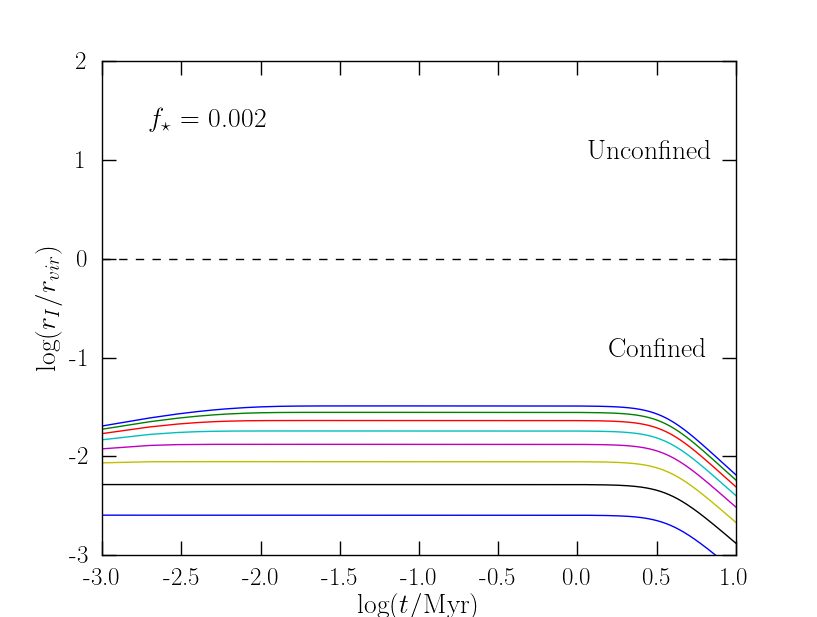}
\caption{Same as Fig. \ref{Fig1}, but for $z=14$.
} 
\label{Fig2}
\end{figure}

We solved numerically equation (\ref{eq10}) for a number of halo
masses with $V=1$ in the range $M=10^{8}$--$10^{11} \msun$ at two
selected redshifts, $z=9$ and $z=14$.  Figures \ref{Fig1} and
\ref{Fig2} show the results for $f_\star=0.2$\%, a reasonable value for
for nearly primordial star-forming halos \citep{Barkana01, Ciardi05, Pawlik12};
note that \cite{Wise09} found that $\langle \fesc f_\star \rangle =0.02$ averaged over all 
atomic cooling ($\Tvir \ge 8000$ K) galaxies assuming a Salpeter IMF. 
%
%
%
%

%
%
IFs in these early halos evolve rapidly to the strong radiative input
by the stars and reach a radius $r_I = \rvir$ within $\sim 1$ Myr,
yielding $\fesc = 0.98-0.90$ for halo masses up to $M\approx 2\times
10^{10}\msun$, beyond which the IF is confined within the halo, and
hence $\fesc=0$. It requires fine tuning for a single object to achieve
intermediate $\fesc$ values; rather, the process favors on-off
states implying that the escape fraction of photons from galaxies can be
intermittent: depending on variations of $f_\star$ in different bursts
(say, due to the increasing gas metallicity or decreasing gas fraction
due to mass loss from supernova-driven winds) the same halo might
switch from being dark to bright above the Lyman limit.  This is consistent
with the results of \cite{Shapley06} who showed that the PDF of the escape fraction
from $z=3$ star-forming galaxies is bimodal (see their Fig. 5).  Note also
that for the largest IF-confined halos, the \HII region starts to
shrink at $t\simgt 3$ Myr as a result of recombinations and the
decline in the supply rate of ionizing photons.

\begin{figure}
\includegraphics[width=90mm]{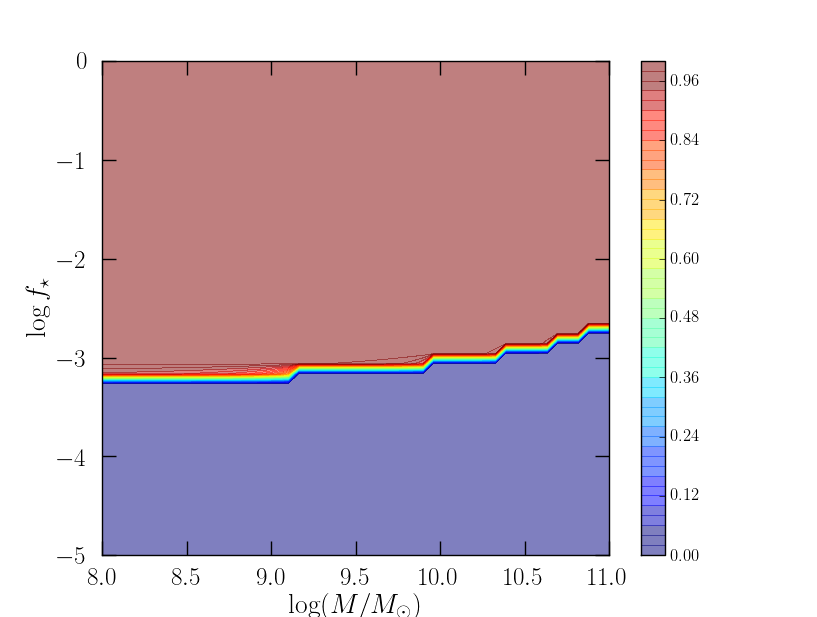}
\caption{Map of $\fesc$ values as a function of halo mass and star
formation efficiency at $z=9$ assuming $V=({\Tvir/T })^{1/2}=1$. The
color scheme defined by the vertical bar on the right hand side.}
\label{Fig3}
\end{figure}

The situation is dramatically different at $z=14$ (for the same
$f_\star$): at that epoch all halos confine their IF and have
$\fesc=0$. This is due to the increase in the gas density and the
larger concentration factor at a fixed mass. A more global view of the
dependence of $\fesc$ on halo mass and star formation efficiency at
$z=9$ is given by Fig. \ref{Fig3}. The abrupt transition from $\fesc
\approx 1$ to zero is very weakly dependent on halo mass, except for
the narrow range of $-2.8 < \log f_\star < -2.2$ in which larger halos
tend to be more opaque to ionizing photons. Above this range, $\fesc =
1$ independently of halo mass and star formation efficiency, and below
this range, $\fesc=0$.
%
%

\begin{figure}
\includegraphics[width=90mm]{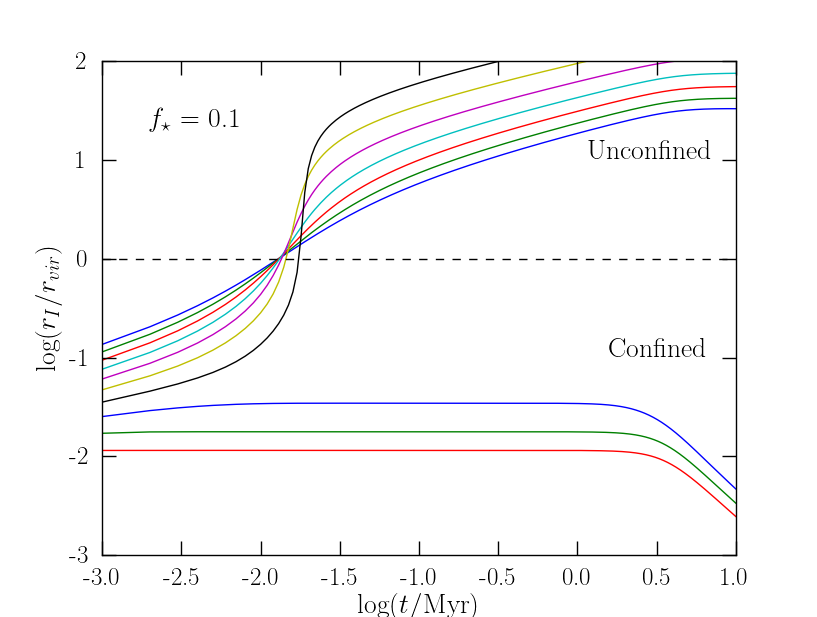}
\caption{Same as Fig. \ref{Fig1}, but for halos with mass in the range
$M=10^{8.+2j/30} \msun$, with $j=0,..,9$ from the top curve ($M=10^8
\msun$) to the bottom one ($M=3\times 10^{8} \msun$). The star
formation efficiency is $f_\star=0.1$ and the gas temperature has been
fixed to $T=2\times 10^4$ K, giving a variable $V$ temperature factor
(see text for details).  }
\label{Fig4}
\end{figure}

Figure \ref{Fig4} shows the examples with $V >1$ introduced in Table
1, for which $T=2 \times 10^4$ K, corresponding to $V^2 = (1.04, 4.82,
22.39, 103.96)$ for halos of mass $M/\msun=(10^8, 10^9, 10^{10},
10^{11})$ and hence resulting in considerably denser and more compact
gas configurations. The IF front is now largely confined within all
halos with $M>10^8 \msun$ even if the star formation efficiency has
been increased by a factor of 50 with respect to the case with $V=1$
shown in Fig.  \ref{Fig1}. In addition, the IF escapes the virial
radius in a very short time\footnote{We have checked that the IF
speed is subluminal, $dr_I/dt < c$}, $t_{out}\sim 10^4$ yr.

\begin{figure}
\includegraphics[width=90mm]{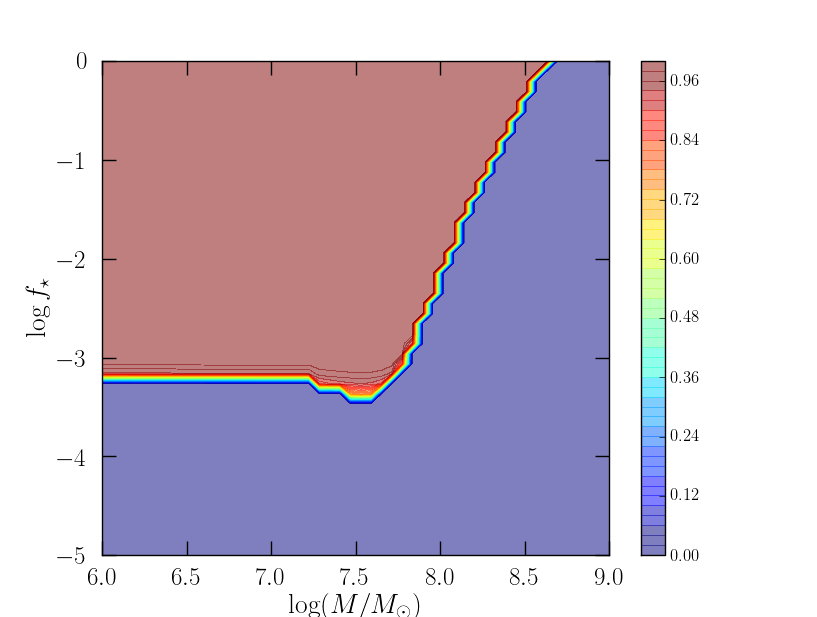}
\caption{Same as Fig. \ref{Fig3} but for a fixed gas temperature
$T=2\times 10^4$ K.  }
\label{Fig5}
\end{figure}

The different behavior relative to the case of $V=1$ is caused by the
associated high gas density (see Table 1). The small scale height of
the gas distribution, correspondingly reduced by a analogous factor $
\propto V^2$, forces the steep acceleration seen just prior to blow
out.  In analogy with the $V=1$ case, we show the dependence of
$\fesc$ on halo mass and star formation efficiency at $z=9$
(Fig. \ref{Fig5}).  In this case the $\fesc=1$ area is much narrower
and limited to small halos with $M< 10^{8.7} \msun$ and $f_\star
\simgt 10^{-3}$.  Thus, according to our model, it is only the
smallest halos that are truly able to inject ionizing photons in the
IGM and therefore substantially contribute to reionization. Larger
halos are instead able to confine their \HII regions very
effectively. At $z=14$ the $\fesc = 1$ region is qualitatively the
same, but its area shrinks because of the larger background density
and the increased concentration parameter at a fixed halo mass.

 To check the effects of the central density we have performed two additional experiments: (a) we have introduced a central density core of size $x_{core} = r_{core}/\rvir$ such that the value the density in the core is equal to $\rho(x < x_{core}) = \rho(x_{core})$; (b) in a more extreme case, we have completely removed the gas, i.e. $\rho(x < x_{core}) = 0$ within the core. The two cases lead to similar conclusions: appreciable deviations from the solutions presented in Fig 4 of the paper are seen only for values of $x_{core} \simgt 0.1$. Stated differently, it would be necessary to completely remove all the gas within 10\% of the virial radius to produce appreciable deviations from what we have obtained with our density profile.

The existence of a minimum value of $f_\star \approx 10^{-3}$ to allow
the escape of photons can be understood based on a simple argument.
This minimum number of ionizing photons provided to each hydrogen atom
in the halo must balance the number of recombinations it undergoes
during the lifetime of the starburst, approximately equal to $t_0 =
10^{6.6}$ yr (see equation \ref{eq8}). This can be expressed by the
following inequality: \ba f_\star &\ge & \frac{C}{{\cal
N}_\gamma}\frac{\alpha^{(2)} t_0}{\mu m_p} (18\pi^2)\Omega_b \rho_c
(z) \\ & = &0.7\times 10^{-3} \left(\frac{C}{3}\right)
\left(\frac{5\times 10^3}{{\cal N}_\gamma}\right)
\left(\frac{1+z}{10}\right)^3 ,
\label{eq11a}
\ea where $C=\langle n_H^2\rangle/\langle n_H\rangle^2\approx 3$ is
the clumping factor accounting for the gas density structure inside a
NFW halo.

The results for halos with $M\simlt 10^8 \msun$ in which $\Tvir \simlt
2 \times 10^4$ K at $z=9$ need some extra attention.  In fact, by
fixing the gas temperature to $2\times 10^4$K, the gas might have
never been accreted in the first place; this is the case if for
example the halo is located within an already ionized region. To
correct this problem, we have set $V=1$ in these mini-halos; of
course, this requires that some cooling agent, such as \HH, HD or LiH,
would cool the gas in these systems to make stars. 
%
%
%
%
%
%
%
%

%
%
\begin{table*}
\begin{minipage}{170mm}
\caption{Values of various quantities related to the global evolution of the mass-averaged escape fraction. See text for more details.}
\begin{center}
\begin{tabular}{|c|c|c|c|c|c|c|c|}
\hline\hline
&\multicolumn{7}{c}{Mass-averaged $\fesc$ --  Constant $f_\star$} \\ 
             & $\Msf$ & $M_{f1}$               & $M_{f0}$      &  $f_{coll}(> \Msf)$   & $f_{coll} (>M_{f0})$   & $\langle\fesc\rangle$  &  $f_\star$  \\ 
            & [$\msun$] &  [$\msun$]  &[$\msun$]  &    &                                    &                \\\hline
$z=6$  &2.977e+08   & 2.977e+08                 &     4.852e+08      &   1.379e-01           &    1.262e-01      & 8.478e-02  & 3.000e-02          \\
$z=9$  & 1.169e+07 & 1.169e+07 & 2.190e+08 & 1.130e-01 & 5.936e-02 & 4.747e-01 & 3.000e-02 \\
$z=9$  & 1.169e+07 & 1.169e+07 & 1.253e+08 & 1.130e-01 & 6.897e-02 & 3.897e-01 & 3.000e-03\\
$z=9$  & 1.169e+07 & 1.169e+07 & 5.818e+08 & 1.130e-01 & 4.374e-02 & 6.130e-01 & 1.000e+00 \\
$z=9$  & 1.169e+07 & 1.169e+07 & 1.169e+07 & 1.130e-01 & 1.130e-01 & 0.000e+00 &  1.000e-04\\
$z=14$ & 1.634e+06 & 1.634e+06 & 8.131e+07 & 5.373e-02 & 1.471e-02 & 7.263e-01 & 3.000e-02 \\\hline\hline
&\multicolumn{7}{c}{Mass-averaged $\fesc$ -- Variable $f_\star(z)$}\\  
             & $\Msf$ & $M_{f1}$               & $M_{f0}$      &  $f_{coll}(> \Msf)$   & $f_{coll} (>M_{f0})$   & $\langle\fesc\rangle$  &  $f_\star$  \\ 
            & [$\msun$] &  [$\msun$]  &[$\msun$]  &    &                                    &                \\\hline
$z=8$   & 5.361e+07 & 5.361e+07 & 2.019e+08 & 1.098e-01 & 8.315e-02 & 2.426e-01 & 8.119e-03\\
$z=10$ & 3.664e+06 & 3.664e+06 & 1.119e+08 & 1.102e-01 & 5.217e-02 & 5.267e-01 & 4.799e-03\\
$z=12$ & 2.030e+06 & 2.030e+06 & 7.128e+07 & 8.003e-02 & 3.148e-02 & 6.066e-01 & 4.281e-03\\\hline
\end{tabular}
\end{center}
\label{Tab2}
\end{minipage}
\end{table*}

\subsection{Global redshift evolution}
Next we would like to constrain the redshift evolution of $\fesc(z)$
when averaged over the entire galaxy population.

The dark matter halo mass function, $n(M,z)$, is well described by the
\cite{Sheth02} form: \be
n(M,z)dM = A\left(1+\frac{1}{\nu'^{2q}}\right) \sqrt{\frac{2}{\pi}}
\frac{\bar\rho}{M} \frac{d\nu'}{dM}
\exp\left(-\frac{\nu'^{2}}{2}\right) dM,
\label{eq12}
\ee with $A=0.322$, $q=0.3$, $\nu'=\sqrt{a}\nu$, $a=0.707$,
$\nu=\delta_c/D(z)\sigma(M)$, with $\delta_c=1.686$, $D(z)$ being the
linear growth factor and $\sigma(M)$ being the r.m.s.  amplitude of
density fluctuations on a mass scale $M$.

We define the mass-averaged escape fraction, $\langle\fesc\rangle$ at
a given $f_\star$ and redshift $z$ as follows: \be
\langle\fesc\rangle(z, f_\star) =1 - \frac {F\left
[>M_{f0}(z,f_\star)\right ]}{F\left[>M_{sf}(z)\right]} ,
\label{eq13}
\ee where $F(>M,z)=\int_M^\infty nM dM/\int_0^\infty nM dM$ is the
collapse fraction of all matter in haloes with mass exceeding $M$ at redshift
$z$.

Equation (\ref{eq13}) contains two important characteristic
masses. The first, $M_{f0}(z,f_\star)$, is defined as the halo mass
above which, for a given redshift $z$ and star formation efficiency
$f_\star$, no ionizing photons can escape, i.e. $\fesc=0$. The results
presented above allow us to precisely determine $M_{f0}(z,f_\star)$.

The second mass scale, $\Msf(z)$, denotes the redshift-dependent
minimum mass of star-forming haloes which is set by radiative
feedback. The evolution of $\Msf$ is determined by two distinct
radiative feedback processes.  The first one is related to the
increase of the cosmological Jeans mass in progressively ionized
cosmic regions; as a result, the infall of gas in haloes below a given
virial temperature is quenched. The evolution of $\Msf(z)$
depends on the details of the reionization history \citep{Gnedin00,
Okamoto08, Schneider08}. Guided by these findings, we adopt the the
value $\Msf=M(\Tvir=T_2)$ with $T_2=3\times 10^4$ K after the end of
reionization, here assumed to occur at $z_{rei}=6$.

Before reionization (at least in the mostly neutral regions) a second
type of feedback, involving the photodissociation of \HH molecules by
the Lyman-Werner (LW) background photons, becomes important. As H$_2$
is the primary coolant for minihalos ($\Tvir< 10^4$ K), fragmentation
of the gas to stars could be suppressed in these objects through H$_2$
dissociation by the UV background \citep{Haiman96, Ciardi00,
Kitayama00, Machacek01}.  According to \cite{Dijkstra04}, $z\sim 10$
mini-halos with $\Tvir = T_1 \approx 2\times 10^3$ K can self-shield
H$_2$ and cool, hence we use this value as the lower threshold for
star formation.  During reionization the interplay between the
photo-ionization heating and LW-feedback is complicated (see the
discussion in \cite{Mesinger08} and Fig. 25 of \cite{Ciardi05}) and
and the evolution of $\Msf$ during this epochs is uncertain. To
circumvent this uncertainty, we follow the phenomenological approach
of \cite{Salvadori09} and \cite{Salvadori12} who suggested that the
metallicity distribution and iron-luminosity relation of the Ultra
Faint dSphs in the Milky Way halo can be explained if these objects
are relic mini-halos that formed their stars prior to
reionization. With this perspective, one can reconstruct $\Msf(z)$ by
fitting local data. This approach yields the functional form, \be \Msf(z) =
M[\Tvir = (T_2-T_1) e^{(z-z_{rei})^2/\Delta z}+T_1] ,
\label{eq14}
\ee with $\Delta z=4$, which we adopt here.

In general $M_{f0}(z,f_\star) > \Msf(z)$. However, for sufficiently
low values of $f_\star$, $M_{f0}(z,f_\star) = \Msf(z)$ and
$\langle\fesc\rangle =0$.  To illustrate this point (see Table
\ref{Tab2}), we examine the results at $z=9$ in Figure \ref{Fig5}.
For $f_\star=0.03$, stars are formed only in halos more massive than
$\Msf = 1.17\times 10^7 \msun$. Ionizing photons can only escape from
halos less massive than $M_{f0} = 2.19\times 10^8 \msun$. As a result,
there is a narrow range of objects that contribute to the ionizing
photons budget in the IGM. Below that range stars cannot form and
above it the gas is sufficiently dense to confine the \HII
region. When weighted with the halo mass function, though, the
collapsed mass fraction above $(\Msf, M_{f0})$ is $(0.113, 0.059)$,
yielding the large value $\langle\fesc\rangle(z=9, f_\star=0.01) =
0.847$. Since \HII regions in abundant low mass halos are unbound, the
average escape fraction is high. The width of the range progressively
shrinks as star formation becomes less efficient.  This is evident
from Table \ref{Tab2}, where for $z=9$ and $f_\star = 10^{-4}$,
$M_{f0}(z,f_\star) = \Msf(z)$ and the escape fraction drops to
zero. These results underline the importance of the possible evolution
in the star formation efficiency, which we address below.
%
%
\begin{figure}
\includegraphics[width=90mm]{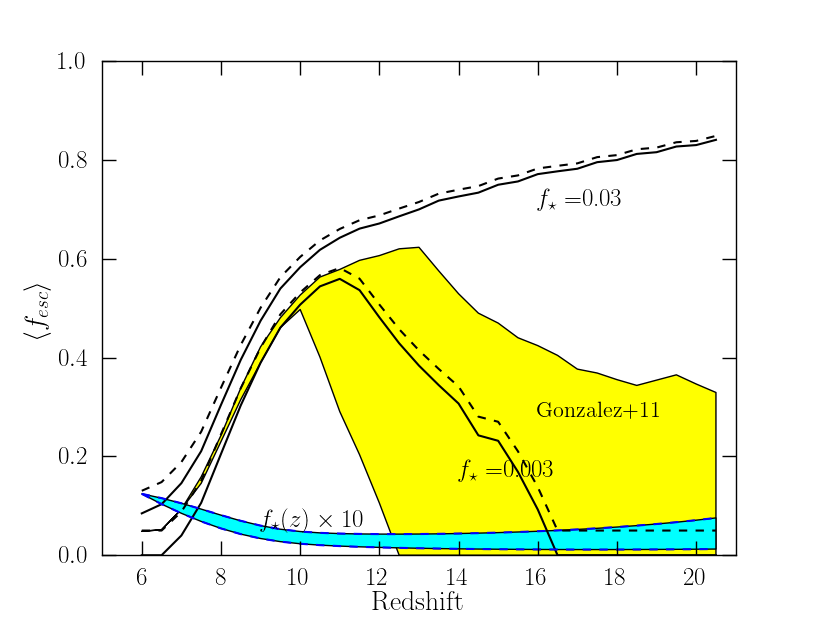}
\caption{Redshift evolution of the mass-averaged escape fraction for
different assumptions concerning the star formation efficiency. Solid
lines refer to the redshift-independent values $f_\star=0.03, 0.003$
as indicated by labels; dotted lines are the same cases in which a fixed 
$\fesc=0.05$ has been assumed for large halos with $M>M_{f0}$ (see Sec. \ref{Con}); the shaded area refers to $f_\star(z)$ (shown
by the shaded cyan area and multiplied by a factor 100) derived from
the stellar mass density evolution by \cite{Gonzalez11} based on equation
(\ref{eq15}).}
\label{Fig6}
\end{figure}

The redshift evolution of the results can be understood in terms of a
balance between two opposite trends. On the one hand, the halo central
densities tend to become larger with increasing redshift, providing a
more efficient confinement of the IF and acting to decrease
$\langle\fesc\rangle$. On the other hand, more dwarf galaxies and
mini-halos are able to produce ionizing photons as the critical mass
$\Msf$ decreases with increasing $z$. The overall evolution for the
two specific values $f_\star = (0.03, 0.003)$ can be seen in Figure
\ref{Fig6} (solid curves). For the high $f_\star$ case, the escape
fraction continues to increase with redshift, reaching
$\langle\fesc\rangle = 0.83$ at $z=20$. If instead $f_\star = 0.003$,
the escape fraction reaches a peak value of $0.59$ at $z=11.5$ where
it starts to decrease until it reaches zero at $z=17.5$ when the
density effect dominates over the enhanced photon production rate.
Interestingly, a nearly redshift-independent $f_\star$ above a
threshold of $\approx 0.1$\% (see Fig. \ref{Fig5}) always leads to
$d\langle\fesc\rangle/dz>0$ during the final phase of reionization.
%
%
\begin{figure}
\includegraphics[width=90mm]{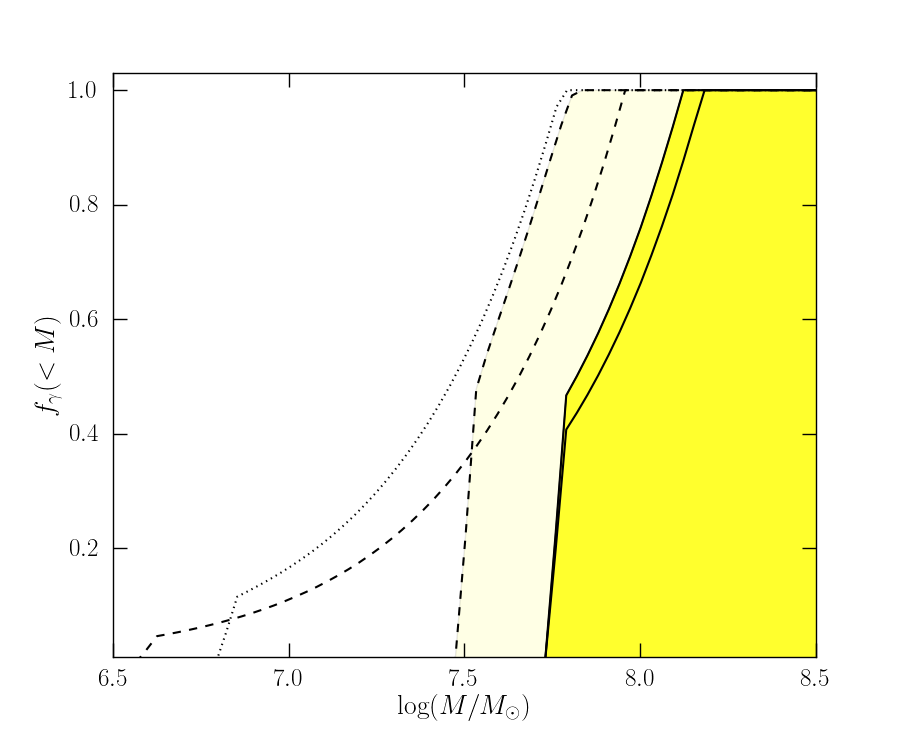}
\caption{Cumulative fraction of ionizing photons produced by halos
below a given halo mass $M$. Lines refer to a constant star formation
efficiency $f_\star=0.003$; shaded regions refer to the upper limit of
the redshift-dependent $f_\star$ (equation \ref{eq15}) based on
\cite{Gonzalez11} SMDs. For both sets of curves, solid (dashed) lines
refer to $z=8$ ($z=10$); for $f_\star=0.003$ the case for $z=12$
(dotted) is also shown.  }
\label{Fig7}
\end{figure}
Although a rather weak mass dependence of $f_\star$ for galaxies of
stellar mass $M_\star > 3\times 10^7 \msun$ is inferred from local
SDSS data \citep{Dekel03}, there are indications of a decreasing star
formation efficiency for high-$z$ galaxies. This is in line also with the 
conclusion (based on numerical simulations) by \cite{Munshi12} that low 
mass ($M < 10^9 M_\odot$) halos have generally lower (0.1\%) values
of $f_\star$. Recently, different
groups \citep{Stark09, Labbe10, Gonzalez11} have obtained stellar
masses of galaxies in the redshift range $4 < z < 7$ from spectral
energy distribution fitting to rest-frame optical and UV fluxes from
Hubble-WFC3/IR camera observations of the Early Release Science field
combined with the deep GOODS-S Spitzer/IRAC data. From this data it
has been possible to reconstruct (see Fig. 4 of \cite{Gonzalez11}) the
evolution of the comoving Stellar Mass Density (SMD), which we define
here as $\rho_\star(z)$. The main result is that $\rho_\star(z=6)=10^7
\msun $ Mpc$^{-3}$; in addition, the SMD grows with decreasing
redshift as $\propto (1 + z)^{-3.4 \pm 0.8}$. We can exploit this
observational result to obtain a robust estimate for $f_\star(z)$
based on the relation, 
\be 
f_\star(z) = \frac{\rho_\star(z)}{F\left[>M_{sf}(z)\right] \Omega_b\rho_c} .
\label{eq15}
\ee 
In other words, we tune the star formation efficiency so that
baryons in collapsed halos account for the abundance of stars observed at
$z<7$. Note that the \cite{Gonzalez11} SMD has also been corrected for
incompleteness at low masses.  The range of allowed values of
$f_\star$ implied by the data and obtained through equation
(\ref{eq15}) are plotted as a shaded area in Fig. \ref{Fig6}. The data
suggest $f_\star = 0.003$ at $z=6$ (for display reasons we have
multiplied the values by a factor 100 in the Figure) decreasing
thereafter by a factor of about three above $z=10$.

The main consequence of a suppressed star formation efficiency in low
mass objects is to introduce a new intermediate mass scale
$M_{f1}$. Because of their lower star formation efficiency, halos above $\Msf$
might not produce enough stars to ensure $\fesc=1$. IFs become unconfined only for
somewhat larger halos of characteristic mass  $M_{f1}$, such that $\Msf <
M_{f1} < M_{f0} $. Thus  $\fesc > 0$ only in the interval
[$M_{f1},M_{f0}$]. Stated differently, even objects that are above $\Msf$ and can
form stars, are now more efficient at confining their IFs; therefore
their escape fraction becomes zero. Table \ref{Tab2} shows how
$M_{f1}$ gradually approaches $M_{f0}$ towards high redshifts, closing
the non-vanishing escape fraction window.

Another interesting question is how the global production of ionizing
photons is distributed among halos of different masses. We quantify
this through the cumulative fraction of ionizing photons produced by
halos below a given mass, \be f_\gamma(< M, z) \propto \int_{M_{sf}}^M
\fesc(M', z, f_\star ) f_\star(\Msf,z ) M' dM',
\label{eq16}
\ee as shown in Figure \ref{Fig7}.  Due to their large escape
fraction, halos with masses $M\simlt 10^8 \msun$ are the dominant
contributors to the cosmic ionizing photon budget. This holds for the
more realistic redshift-dependent $f_\star$, but also for the constant
$f_\star$ case. Thus, the low-mass galaxies and mini-halos might have
dominated reionization. In particular, at $z=10$ the mass
corresponding to $T=10^4$ K limiting the mini-halo mass range from
above, is $\sim 3\times 10^7 \msun$. Figure \ref{Fig7} shows that
independently of the star formation efficiency prescription, we
predict that mini-halos contribute $\sim 40\%$ of the total
emissivity.

Reionization models suggest an increasing emissivity towards high
redshifts \citep{Bolton07, Mitra12a, Mitra12b, Kuhlen12}, whose most
natural explanation is in terms of a $\fesc$ increasing with redshift
as we find here. For example, \cite{Mitra12b} find that: {\it (i)} the
escape fraction increases from $\langle\fesc\rangle =
0.068_{-0.047}^{+0.054}$ at z=6 to $0.179_{-0.132}^{+0.331}$ at $z=8$;
and {\it (ii)} at $z=10$ $\langle\fesc\rangle > 0.146$.


\section{Discussion}
\label{Con}

We have calculated the escape fraction of ionizing photons from
starburst galaxies during reionization, assuming that galactic gas
cools down to a temperature floor ($\sim 10^4$K for atomic hydrogen
and $\sim 2\times 10^3$K for self-shielded molecular hydrogen).  We
have found that most of the escaping ionizing photons originates in
the lowest mass galaxies with $T_{\rm vir} \simlt 10^4~{\rm K}$.  Due
to the shallowness of the gravitational-potential well in these
galactic halos, the gas maintains a low density and its recombination
rate is low. This allows the ionization front to break through the
virial radius rapidly, resulting in a high escape fraction of the
emitted ionizing radiation. As reionization progresses, the
intergalactic medium is photo-heated and the abundance of galaxies
with $T_{\rm vir}\sim 10^4$K declines, leading to a decrease in
$\langle f_{\rm esc}\rangle$.  This physical mechanism may account for
the increase of $\langle f_{\rm esc}\rangle$ with redshift inferred by
\cite{Mitra12b}.

Our study is consistent with previous studies of low mass galaxies by
\cite{Wise09}, who found a high escape fraction ($> 0.1$). Low mass
galaxies do not form a disk \citep{Pawlik11}, which could trap
effectively ionizing photons. The relative importance of minihalos for
the global production rate of ionizing photons is determined by the
product $\fesc \times f_\star$. Here we have used a phenomenological
calibration of $f_\star$. In order to produce the observed stellar
density at $z=6$--$7$ by minihalos, $f_\star$ has to be $\simgt
0.001$. This value lies in the middle range resulting from
\cite{Wise09} idealized halo simulations, but they find the star
formation efficiency in halos with $M< 10^7 \msun$ to be smaller.  We
find that the required increase of $\fesc$ with redshift cannot be
achieved if minihalos are sterile or form stars inefficiently.

Our simplified calculation ignored angular momentum which drives cold
gas to a rotating disk configuration. Our model allowed cold gas to
condense to arbitrarily small scales without making a disk, and so one
might worry that it underestimates $f_{\rm esc}$. However, previous
calculations have demonstrated that on the scale of galactic disks the
escape fraction is very small \citep{Wood00,Gnedin08, Razoumov06,
Wise09}, and so our inclusion of angular momentum could not increase
$f_{\rm esc}$ significantly. In particular, the conclusion that
$f_{\rm esc}$ is high in low-mass halos must be robust.  Although our
simplified model predicts zero escape fractions from brief starbursts
in massive halos, the actual value of $f_{\rm esc}$ would be finite in
reality due to persistent star formation histories and winds driven by
supernova feedback or Ly$\alpha$ radiation pressure
\citep{Dijkstra08}.  Our results are in qualitative agreement with
\cite{Gnedin08} and \cite{Razoumov06} who simulated more massive
galaxies ($M \simgt 10^{10} \msun$) and inferred an extremely small
escape fraction values $\fesc\simlt 1$\%), even in situations where
the star formation rate is high ($\sim 10 \msun$ yr$^{-1}$). This
results from the high gas densities in massive galaxies due the
presence of a rotationally-supported, geometrically-thin disk.

To get a simple estimate of the impact of these larger galaxies, 
we have also run cases allowing a certain ionizing photon fraction
(here assumed to be 5\%, guided by local observations) to escape from
halos of mass $M > M_{f0}$. 
The results are shown for two cases in Fig. 6 (dotted lines). 
The reported trend of an increasing $\fesc$ for $z<11$ remains clearly visible.

Finally, we note that the trend we find for the increase of the escape
fraction with redshift should be enhanced if the initial mass function
of stars was tilted towards more massive stars. This is because a
higher production rate of ionizing photons would allow the
ionization front to ionize the surrounding hydrogen blanket even
faster than we calculated.

\section*{Acknowledgments} 
Our computations were based on the Scipy open source scientific tools
for Python \texttt{http://www.scipy.org/}. This work was supported in
part by NSF grant AST-0907890 and NASA grants NNX08AL43G and
NNA09DB30A.

\bibliographystyle{apj}
\bibliography{ref}

\begin{thebibliography}{58}
\expandafter\ifx\csname natexlab\endcsname\relax\def\natexlab#1{#1}\fi

\bibitem[{{Barkana} \& {Loeb}(2001)}]{Barkana01}
{Barkana}, R., \& {Loeb}, A. 2001, \physrep, 349, 125

\bibitem[{{Bolton} \& {Haehnelt}(2007)}]{Bolton07}
{Bolton}, J.~S., \& {Haehnelt}, M.~G. 2007, \mnras, 382, 325

\bibitem[{{Boutsia} {et~al.}(2011){Boutsia}, {Grazian}, {Giallongo}, {Fontana},
  {Pentericci}, {Castellano}, {Zamorani}, {Mignoli}, {Vanzella}, {Fiore},
  {Lilly}, {Gallozzi}, {Testa}, {Paris}, \& {Santini}}]{Boutsia11}
{Boutsia}, K., {et~al.} 2011, \apj, 736, 41

\bibitem[{{Bouwens} {et~al.}(2012){Bouwens}, {Illingworth}, {Oesch}, {Trenti},
  {Labb{\'e}}, {Franx}, {Stiavelli}, {Carollo}, {van Dokkum}, \&
  {Magee}}]{Bouwens12}
{Bouwens}, R.~J., {et~al.} 2012, \apjl, 752, L5

\bibitem[{{Caffau} {et~al.}(2011){Caffau}, {Bonifacio}, {Fran{\c c}ois},
  {Sbordone}, {Monaco}, {Spite}, {Spite}, {Ludwig}, {Cayrel}, {Zaggia},
  {Hammer}, {Randich}, {Molaro}, \& {Hill}}]{Caffau11}
{Caffau}, E., {et~al.} 2011, \nat, 477, 67

\bibitem[{{Cayrel} {et~al.}(2004){Cayrel}, {Depagne}, {Spite}, {Hill}, {Spite},
  {Fran{\c c}ois}, {Plez}, {Beers}, {Primas}, {Andersen}, {Barbuy},
  {Bonifacio}, {Molaro}, \& {Nordstr{\"o}m}}]{Cayrel04}
{Cayrel}, R., {et~al.} 2004, \aap, 416, 1117

\bibitem[{{Ciardi} \& {Ferrara}(2005)}]{Ciardi05}
{Ciardi}, B., \& {Ferrara}, A. 2005, \ssr, 116, 625

\bibitem[{{Ciardi} {et~al.}(2000){Ciardi}, {Ferrara}, \& {Abel}}]{Ciardi00}
{Ciardi}, B., {Ferrara}, A., \& {Abel}, T. 2000, \apj, 533, 594

\bibitem[{{Dekel} \& {Woo}(2003)}]{Dekel03}
{Dekel}, A., \& {Woo}, J. 2003, \mnras, 344, 1131

\bibitem[{{Dijkstra} {et~al.}(2004){Dijkstra}, {Haiman}, {Rees}, \&
  {Weinberg}}]{Dijkstra04}
{Dijkstra}, M., {Haiman}, Z., {Rees}, M.~J., \& {Weinberg}, D.~H. 2004, \apj,
  601, 666

\bibitem[{{Dijkstra} \& {Loeb}(2008)}]{Dijkstra08}
{Dijkstra}, M., \& {Loeb}, A. 2008, \mnras, 391, 457

\bibitem[{{Dove} {et~al.}(2000){Dove}, {Shull}, \& {Ferrara}}]{Dove00}
{Dove}, J.~B., {Shull}, J.~M., \& {Ferrara}, A. 2000, \apj, 531, 846

\bibitem[{{Fern{\'a}ndez-Soto} {et~al.}(2003){Fern{\'a}ndez-Soto}, {Lanzetta},
  \& {Chen}}]{Fernandez03}
{Fern{\'a}ndez-Soto}, A., {Lanzetta}, K.~M., \& {Chen}, H.-W. 2003, \mnras,
  342, 1215

\bibitem[{{Finkelstein} {et~al.}(2012){Finkelstein}, {Papovich}, {Ryan},
  {Pawlik}, {Dickinson}, {Ferguson}, {Finlator}, {Koekemoer}, {Giavalisco},
  {Cooray}, {Dunlop}, {Faber}, {Grogin}, {Kocevski}, \&
  {Newman}}]{Finkelstein12}
{Finkelstein}, S.~L., {et~al.} 2012, ArXiv e-prints

\bibitem[{{Giallongo} {et~al.}(2008){Giallongo}, {Ragazzoni}, {Grazian},
  {Baruffolo}, {Beccari}, {de Santis}, {Diolaiti}, {di Paola}, {Farinato},
  {Fontana}, {Gallozzi}, {Gasparo}, {Gentile}, {Green}, {Hill}, {Kuhn},
  {Pasian}, {Pedichini}, {Radovich}, {Salinari}, {Smareglia}, {Speziali},
  {Testa}, {Thompson}, {Vernet}, \& {Wagner}}]{Giallongo08}
{Giallongo}, E., {et~al.} 2008, \aap, 482, 349

\bibitem[{{Gnedin}(2000)}]{Gnedin00}
{Gnedin}, N.~Y. 2000, \apj, 542, 535

\bibitem[{{Gnedin} {et~al.}(2008){Gnedin}, {Kravtsov}, \& {Chen}}]{Gnedin08}
{Gnedin}, N.~Y., {Kravtsov}, A.~V., \& {Chen}, H.-W. 2008, \apj, 672, 765

\bibitem[{{Gonz{\'a}lez} {et~al.}(2011){Gonz{\'a}lez}, {Labb{\'e}}, {Bouwens},
  {Illingworth}, {Franx}, \& {Kriek}}]{Gonzalez11}
{Gonz{\'a}lez}, V., {Labb{\'e}}, I., {Bouwens}, R.~J., {Illingworth}, G.,
  {Franx}, M., \& {Kriek}, M. 2011, \apjl, 735, L34

\bibitem[{{Grazian} {et~al.}(2011){Grazian}, {Castellano}, {Koekemoer},
  {Fontana}, {Pentericci}, {Testa}, {Boutsia}, {Giallongo}, {Giavalisco}, \&
  {Santini}}]{Grazian11}
{Grazian}, A., {et~al.} 2011, \aap, 532, A33

\bibitem[{{Greif} {et~al.}(2012){Greif}, {Bromm}, {Clark}, {Glover}, {Smith},
  {Klessen}, {Yoshida}, \& {Springel}}]{Greif12}
{Greif}, T.~H., {Bromm}, V., {Clark}, P.~C., {Glover}, S.~C.~O., {Smith},
  R.~J., {Klessen}, R.~S., {Yoshida}, N., \& {Springel}, V. 2012, \mnras, 424,
  399

\bibitem[{{Haiman} {et~al.}(1996){Haiman}, {Rees}, \& {Loeb}}]{Haiman96}
{Haiman}, Z., {Rees}, M.~J., \& {Loeb}, A. 1996, \apj, 467, 522

\bibitem[{{Inoue} {et~al.}(2006){Inoue}, {Iwata}, \& {Deharveng}}]{Inoue06}
{Inoue}, A.~K., {Iwata}, I., \& {Deharveng}, J.-M. 2006, \mnras, 371, L1

\bibitem[{{Iwata} {et~al.}(2009){Iwata}, {Inoue}, {Matsuda}, {Furusawa},
  {Hayashino}, {Kousai}, {Akiyama}, {Yamada}, {Burgarella}, \&
  {Deharveng}}]{Iwata09}
{Iwata}, I., {et~al.} 2009, \apj, 692, 1287

\bibitem[{{Kitayama} {et~al.}(2000){Kitayama}, {Tajiri}, {Umemura}, {Susa}, \&
  {Ikeuchi}}]{Kitayama00}
{Kitayama}, T., {Tajiri}, Y., {Umemura}, M., {Susa}, H., \& {Ikeuchi}, S. 2000,
  \mnras, 315, L1

\bibitem[{{Kuhlen} \& {Faucher-Gigu{\`e}re}(2012)}]{Kuhlen12}
{Kuhlen}, M., \& {Faucher-Gigu{\`e}re}, C.-A. 2012, \mnras, 423, 862

\bibitem[{{Labb{\'e}} {et~al.}(2010){Labb{\'e}}, {Gonz{\'a}lez}, {Bouwens},
  {Illingworth}, {Oesch}, {van Dokkum}, {Carollo}, {Franx}, {Stiavelli},
  {Trenti}, {Magee}, \& {Kriek}}]{Labbe10}
{Labb{\'e}}, I., {et~al.} 2010, \apjl, 708, L26

\bibitem[{{Larson} {et~al.}(2011){Larson}, {Dunkley}, {Hinshaw}, {Komatsu},
  {Nolta}, {Bennett}, {Gold}, {Halpern}, {Hill}, {Jarosik}, {Kogut}, {Limon},
  {Meyer}, {Odegard}, {Page}, {Smith}, {Spergel}, {Tucker}, {Weiland},
  {Wollack}, \& {Wright}}]{Larson11}
{Larson}, D., {et~al.} 2011, \apjs, 192, 16

\bibitem[{{Leitherer} {et~al.}(1999){Leitherer}, {Schaerer}, {Goldader},
  {Gonz{\'a}lez Delgado}, {Robert}, {Kune}, {de Mello}, {Devost}, \&
  {Heckman}}]{Leitherer99}
{Leitherer}, C., {et~al.} 1999, \apjs, 123, 3

\bibitem[{{Loeb} \& {Furlanetto}(2013)}]{Loeb13}
{Loeb}, A., \& {Furlanetto}, S. 2013, {The First Galaxies in the Universe}
  (Princeton University Press, in press)

\bibitem[{{Machacek} {et~al.}(2001){Machacek}, {Bryan}, \& {Abel}}]{Machacek01}
{Machacek}, M.~E., {Bryan}, G.~L., \& {Abel}, T. 2001, \apj, 548, 509

\bibitem[{{Makino} {et~al.}(1998){Makino}, {Sasaki}, \& {Suto}}]{Makino98}
{Makino}, N., {Sasaki}, S., \& {Suto}, Y. 1998, \apj, 497, 555

\bibitem[{{Maselli} {et~al.}(2003){Maselli}, {Ferrara}, \&
  {Ciardi}}]{Maselli03}
{Maselli}, A., {Ferrara}, A., \& {Ciardi}, B. 2003, \mnras, 345, 379

\bibitem[{{Mesinger} \& {Dijkstra}(2008)}]{Mesinger08}
{Mesinger}, A., \& {Dijkstra}, M. 2008, \mnras, 390, 1071

\bibitem[{{Mitra} {et~al.}(2012{\natexlab{a}}){Mitra}, {Choudhury}, \&
  {Ferrara}}]{Mitra12a}
{Mitra}, S., {Choudhury}, T.~R., \& {Ferrara}, A. 2012{\natexlab{a}}, \mnras,
  419, 1480

\bibitem[{{Mitra} {et~al.}(2012{\natexlab{b}}){Mitra}, {Ferrara}, \&
  {Choudhury}}]{Mitra12b}
{Mitra}, S., {Ferrara}, A., \& {Choudhury}, T.~R. 2012{\natexlab{b}}, ArXiv
  e-prints

\bibitem[{{Mu{\~n}oz} \& {Loeb}(2011)}]{Munoz11}
{Mu{\~n}oz}, J.~A., \& {Loeb}, A. 2011, \apj, 729, 99

\bibitem[{{Munshi} {et~al.}(2012){Munshi}, {Governato}, {Brooks},
  {Christensen}, {Shen}, {Loebman}, {Moster}, {Quinn}, \& {Wadsley}}]{Munshi12}
{Munshi}, F., {et~al.} 2012, ArXiv e-prints

\bibitem[{{Navarro} {et~al.}(1997){Navarro}, {Frenk}, \& {White}}]{Navarro97}
{Navarro}, J.~F., {Frenk}, C.~S., \& {White}, S.~D.~M. 1997, \apj, 490, 493

\bibitem[{{Okamoto} {et~al.}(2008){Okamoto}, {Gao}, \& {Theuns}}]{Okamoto08}
{Okamoto}, T., {Gao}, L., \& {Theuns}, T. 2008, \mnras, 390, 920

\bibitem[{{Pawlik} {et~al.}(2011){Pawlik}, {Milosavljevi{\'c}}, \&
  {Bromm}}]{Pawlik11}
{Pawlik}, A.~H., {Milosavljevi{\'c}}, M., \& {Bromm}, V. 2011, \apj, 731, 54

\bibitem[{{Pawlik} {et~al.}(2012){Pawlik}, {Milosavljevic}, \&
  {Bromm}}]{Pawlik12}
{Pawlik}, A.~H., {Milosavljevic}, M., \& {Bromm}, V. 2012, ArXiv e-prints

\bibitem[{{Prada} {et~al.}(2012){Prada}, {Klypin}, {Cuesta}, {Betancort-Rijo},
  \& {Primack}}]{Prada12}
{Prada}, F., {Klypin}, A.~A., {Cuesta}, A.~J., {Betancort-Rijo}, J.~E., \&
  {Primack}, J. 2012, \mnras, 423, 3018

\bibitem[{{Razoumov} \& {Sommer-Larsen}(2006)}]{Razoumov06}
{Razoumov}, A.~O., \& {Sommer-Larsen}, J. 2006, \apjl, 651, L89

\bibitem[{{Salvadori} \& {Ferrara}(2009)}]{Salvadori09}
{Salvadori}, S., \& {Ferrara}, A. 2009, \mnras, 395, L6

\bibitem[{{Salvadori} \& {Ferrara}(2012)}]{Salvadori12}
---. 2012, \mnras, 421, L29

\bibitem[{{Salvaterra} {et~al.}(2011){Salvaterra}, {Ferrara}, \&
  {Dayal}}]{Salvaterra11}
{Salvaterra}, R., {Ferrara}, A., \& {Dayal}, P. 2011, \mnras, 414, 847

\bibitem[{{Schneider} {et~al.}(2008){Schneider}, {Salvaterra}, {Choudhury},
  {Ferrara}, {Burigana}, \& {Popa}}]{Schneider08}
{Schneider}, R., {Salvaterra}, R., {Choudhury}, T.~R., {Ferrara}, A.,
  {Burigana}, C., \& {Popa}, L.~A. 2008, \mnras, 384, 1525

\bibitem[{{Shapley} {et~al.}(2006){Shapley}, {Steidel}, {Pettini},
  {Adelberger}, \& {Erb}}]{Shapley06}
{Shapley}, A.~E., {Steidel}, C.~C., {Pettini}, M., {Adelberger}, K.~L., \&
  {Erb}, D.~K. 2006, \apj, 651, 688

\bibitem[{{Sheth} \& {Tormen}(2002)}]{Sheth02}
{Sheth}, R.~K., \& {Tormen}, G. 2002, \mnras, 329, 61

\bibitem[{{Siana} {et~al.}(2007){Siana}, {Teplitz}, {Colbert}, {Ferguson},
  {Dickinson}, {Brown}, {Conselice}, {de Mello}, {Gardner}, {Giavalisco}, \&
  {Menanteau}}]{Siana07}
{Siana}, B., {et~al.} 2007, \apj, 668, 62

\bibitem[{{Stark} {et~al.}(2009){Stark}, {Ellis}, {Bunker}, {Bundy}, {Targett},
  {Benson}, \& {Lacy}}]{Stark09}
{Stark}, D.~P., {Ellis}, R.~S., {Bunker}, A., {Bundy}, K., {Targett}, T.,
  {Benson}, A., \& {Lacy}, M. 2009, \apj, 697, 1493

\bibitem[{{Steidel} {et~al.}(2001){Steidel}, {Pettini}, \&
  {Adelberger}}]{Steidel01}
{Steidel}, C.~C., {Pettini}, M., \& {Adelberger}, K.~L. 2001, \apj, 546, 665

\bibitem[{{Trac} \& {Cen}(2007)}]{Trac07}
{Trac}, H., \& {Cen}, R. 2007, \apj, 671, 1

\bibitem[{{Vanzella} {et~al.}(2012{\natexlab{a}}){Vanzella}, {Guo},
  {Giavalisco}, {Grazian}, {Castellano}, {Cristiani}, {Dickinson}, {Fontana},
  {Nonino}, {Giallongo}, {Pentericci}, {Galametz}, {Faber}, {Ferguson},
  {Grogin}, {Koekemoer}, {Newman}, \& {Siana}}]{Vanzella12a}
{Vanzella}, E., {et~al.} 2012{\natexlab{a}}, \apj, 751, 70

\bibitem[{{Vanzella} {et~al.}(2012{\natexlab{b}}){Vanzella}, {Nonino},
  {Cristiani}, {Rosati}, {Zitrin}, {Bartelmann}, {Grazian}, {Broadhurst},
  {Meneghetti}, \& {Grillo}}]{Vanzella12b}
---. 2012{\natexlab{b}}, \mnras, 424, L54

\bibitem[{{Wise} \& {Cen}(2009)}]{Wise09}
{Wise}, J.~H., \& {Cen}, R. 2009, \apj, 693, 984

\bibitem[{{Wood} \& {Loeb}(2000)}]{Wood00}
{Wood}, K., \& {Loeb}, A. 2000, \apj, 545, 86

\bibitem[{{Wyithe} \& {Loeb}(2003)}]{Wyithe03}
{Wyithe}, J.~S.~B., \& {Loeb}, A. 2003, \apj, 586, 693

\end{thebibliography}

\newpage 
\label{lastpage} 
\end{document}